\documentclass[aps,prb,reprint, superscriptaddress]{revtex4-2}

\usepackage[version=4]{mhchem}
\usepackage{caption}
\usepackage{subscript}
\usepackage{graphicx}
\usepackage{comment}
\usepackage{subcaption}
\usepackage{float}
\usepackage{ragged2e}

\begin{document}
\raggedbottom


\title{Field-Induced Ferroelectric Phase Transition Dynamics in PMN-PT compositions  near the Morphotropic Phase Boundary}


\author{Shivjeet Chanan}

\author{Joseph A. Kerchenfaut}

\author{Eduard Ilin}

\author{Eugene V. Colla}
\email{kolla@illinois.edu}
\affiliation{Department of Physics, University of Illinois at Urbana-Champaign, 1110 W Green St Urbana IL 61801-3080, USA}


\date{\today}

\begin{abstract}
The dynamical behavior of field-induced ferroelectric phase transitions in compositions of \ce{PbMg_{1/3}Nb_{2/3}O3}(1-x)-\ce{PbTiO3}(x), called PMN-PT, near the Morphotropic Phase Boundary (MPB) was investigated through several different external electric field application protocols. Our results show that PMN‑PT compositions near the MPB exhibit phase‑transition dynamics that differ markedly from those of compositions far below the MPB. We demonstrate that the electric-field history has a notable impact on the field-induced transition temperature $T_C$, Zero-Field Cooling (ZFC) delay time $\tau_{ZFC}$ , and induced polarization $P_c$, gained/lost in the transition. Furthermore, we show that under specific field–temperature conditions, PMN‑PT can retain a memory of its electric‑field history and use it to kinetically accelerate ferroelectric ordering. An explanation for the key differences in phase‑transition dynamics between MPB‑proximal and MPB‑distant compositions is proposed and contextualized within prior literature.
\end{abstract}


\maketitle

\section{Introduction}

\begin{figure}[h!]
    \captionsetup{width=\linewidth, format=plain, justification=justified}
    \centering
    \includegraphics[width=0.475\textwidth]{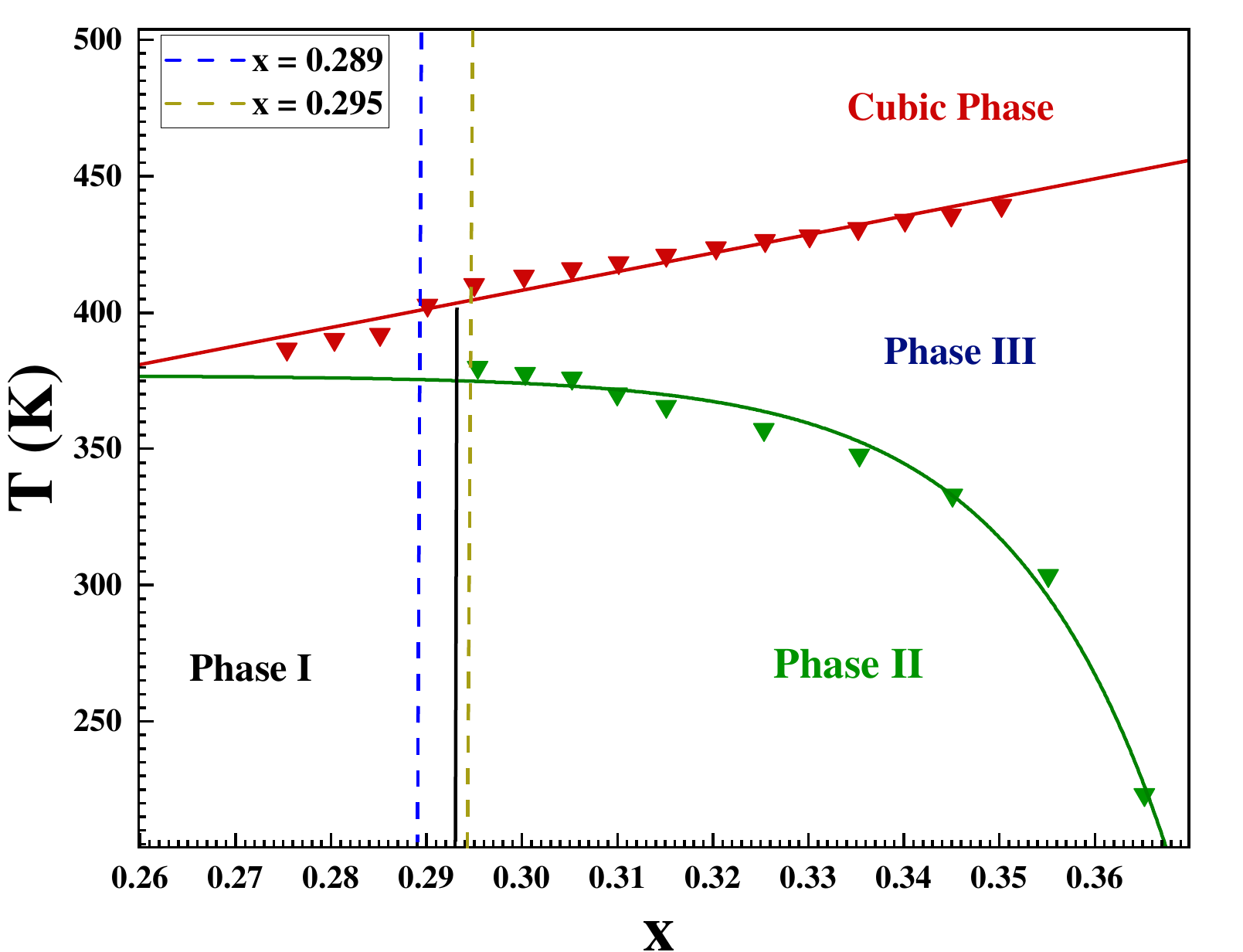}
    \caption{\justifying 
    An empirical concentration-temperature phase diagram for PMN-PT was adopted from \cite{Zekria05}. Phase II represents many possible ferroelectric phases ranging from orthorhombic to monoclinic that lie within the morphotropic phase boundary for PMN-PT. Phase I typically represents the Pseudo-cubic/Rhombohedral, while Phase III typically represents the Tetragonal ferroelectric state. On the phase diagram, the colored dashed lines represent the concentration of each sample that was studied.}
    \label{fig:figure1}
\end{figure}

PMN-PT is a condensed matter system that has multiple parameters affecting its macroscopic phase such as external electric fields, temperature, pressure, and the concentration of \ce{PbTiO_3} \cite{Wu2017, Noheda2002, Ahart2012, Kim2022}. Two phase diagrams can be constructed from these parameters: Temperature ($T$) versus Concentration of \ce{PbTiO_3} ($x$)  and Electric Field ($E$) versus Temperature for a fixed concentration, $x$ \cite{Davis2006}. Fig.~\ref{fig:figure1} shows an adapted $T$-$x$ phase diagram of PMN-PT which indicates the various types of induced ferroelectric states, distinguished by different crystalline lattice structures \cite{Guo2003}. A region called the Morphotropic Phase Boundary (MPB) delineates different ferroelectric states by their crystallographic structure \cite{Guo2003}. In Fig.~\ref{fig:figure1}, this is represented in the Phase II region. Prior literature has established that different ferroelectric modes in this region not only compete but also coexist in this region \cite{Zhang2017}. Thus, a PMN-PT crystal with a concentration $x$ near the MPB is an excellent system to study the effects on macroscopic properties due to the competition and coexistence of multiple types of long-range ferroelectric order.

Yet, this is not the only length scale at which polar order exists within PMN-PT. PMN-PT is an engineered material where a regular relaxor, \ce{Pb(Mg_{1/3}Nb_{2/3})O3}, is doped with a concentration, $x$, of a regular ferroelectric, \ce{PbTiO_3}. Although there is no single universally accepted definition of a relaxor, a typical relaxor exhibits three key properties. First, its dielectric response is diffuse, rounded, and dispersive in nature \cite{Cross1993}. Second, there is no macroscopic change in its crystallographic structure when the system gains relaxor properties \cite{Cross1993}. Third, the dielectric susceptibility does not follow the Curie-Weiss Law relation, $\chi = \frac{C}{T - T_c}$, as standard ferroelectrics do \cite{Cross1993}. These properties can be explained by the formation of local, short-range polar order, known as Polar Nanoregions (PNRs), attributed to the off-center displacement of atoms within the unit cell \cite{Burns1986}. These local, short-range polar structures lead to glassy behavior, where they are frozen in place below a freezing-temperature, $T_f$ (derived from Vogel-Fulcher relations) \cite{Levstik1998, Delgado2009}.

The nature of field-induced ferroelectric phase transitions in relaxor ferroelectrics can be altered by varying parameters of the experimental protocols used to generate the phase transition. For example, in compositions of PMN-PT far from the MPB ($x \sim 0.12$), as the cooling rate is reduced, the temperature at which field-cooled ferroelectricity is reached increases \cite{Colla2007}, as shown in  Fig.~\ref{fig:figure3} (a). The electric field history is another key factor that significantly affects features of field-induced phase transitions in relaxor ferroelectrics. In compositions near the MPB of \ce{PbZn_{1/3}Nb_{2/3}O_3}-\ce{PbTiO_3} (PZN-PT), it was shown that a prior field-cooled phase transition could be remembered in a subsequent zero-field cooling phase transition \cite{Xu2005, Farnsworth2011, Wang2016}. Similar memory effects were observed in other relaxor-ferroelectrics including \ce{Sr_{0.61}Ba_{0.39}Nb_2O_6} doped with cerium (SBN-Cerium) \cite{Granzow2002}. Further neutron diffuse scattering measurements on PZN-PT revealed that under application of a moderately strong electric field, short-range polar structures coexisted with macroscopic ferroelectric domains \cite{Xu2006}.

Though studied in other relaxor-ferroelectric systems, the field-induced ferroelectric phase transition dynamics in compositions of PMN-PT near its MPB are not well understood. To the extent of the authors' knowledge, there is currently an insufficient amount of literature that seeks to determine if the kinetics and dynamics of such phase transitions significantly differ in compositions of PMN-PT near and far from its MPB. In this paper, we will show that the behavior exhibited by field-induced ferroelectric phase transitions in compositions of PMN-PT near the MPB differ notably from the behavior exhibited in compositions of PMN-PT far from the MPB. Our principal results are three-fold. First, our results demonstrate that the electric-field, temperature history dependence for PMN-PT compositions near and far from the MPB fundamentally differ. Second, our results suggest that the phase-transition dynamics exhibit kinetic dampening when aged in certain temperature regions, translating to notably delayed field-induced ferroelectric phase transitions. Third, our results show that this dampening behavior can be ``overridden" under repetitive electric field-temperature cycling. Additionally, we find an exotic ``memory" effect, where PMN-PT compositions near the MPB can undergo ferroelectric phase transitions without requiring concurrent application of an electric field. This occurs only when a certain electric field-temperature history was induced on the composition.

\section{Sample Preparation and Experimental Methodology}
In our experiments, we used two samples, with [111] orientations, where $x = 0.289, 0.295$. They are labeled by their respective concentrations. Each sample was grown by MSE Supplies using the Bridgman technique (akin to the sample used in \cite{Colla2007}). Each PMN-PT sample was configured as a capacitor. A 100 nm Ag thick film was layered on top of 10 nm of Cr, prepared through in vacuum thin-film deposition to act as electrode plates deposited on top of PMN-PT samples. The electrode areas were approximately 6.0 mm\textsuperscript{2}, and thickness of the samples were approximately 0.5 mm. The sample-based capacitor system was mounted onto a measurement probe in a sealed cryostat chamber. AC and DC electric fields were applied to the sample. The AC current response was measured by SR830 lock-in amplifier, which provided information on the dielectric susceptibility. The DC component of the current was measured by a digital voltmeter. The variation of the polarization was calculated by integrating the polarization current.

\begin{figure}[t!]
    \centering
    \includegraphics[width=0.485\textwidth]{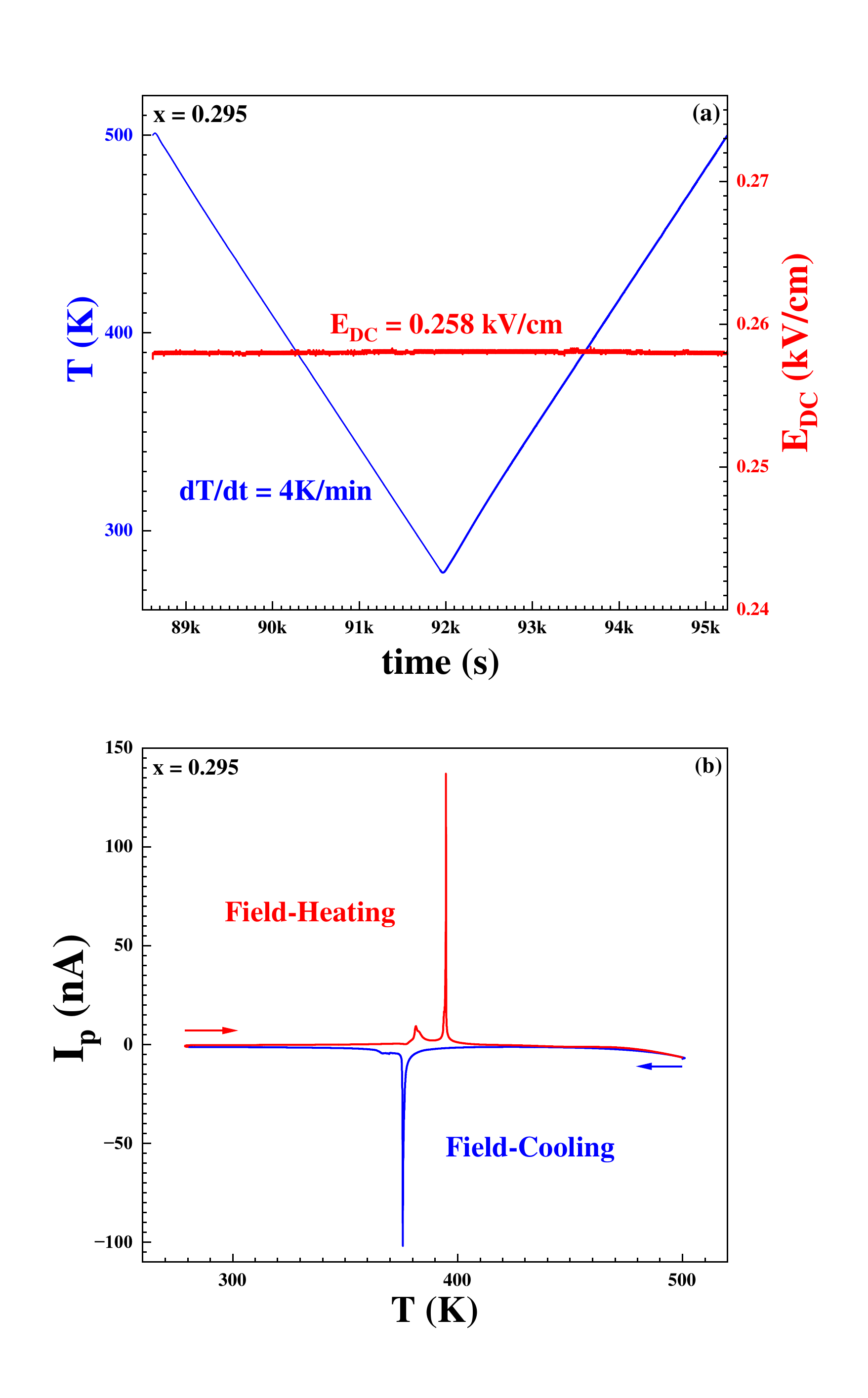}
    \caption{\justifying (a) FC-FH Regime Protocol; (b) Polarization current responses during FC and FH.}
    \label{fig:figure2}
\end{figure} 

\begin{figure*}[htbp!]
    \centering
    \includegraphics[width=1.0\textwidth]{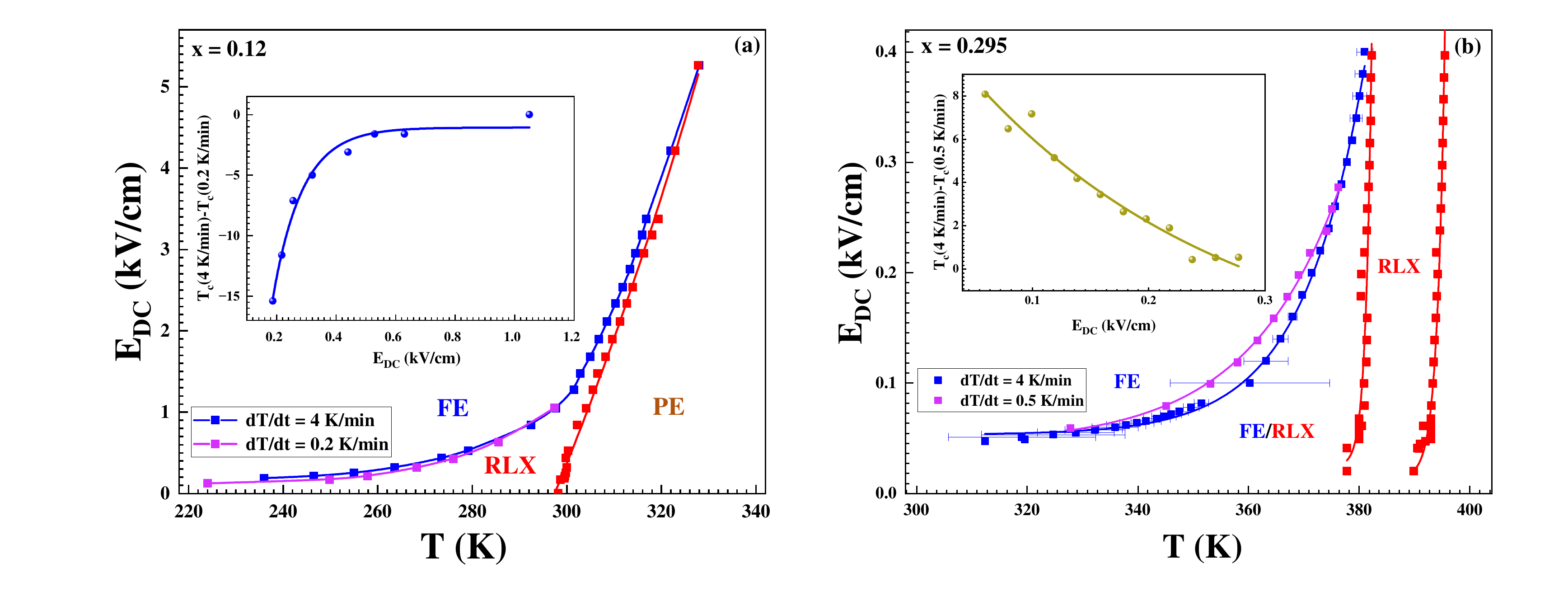}
    \caption{\justifying Empirical field-induced electric field-temperature phase diagrams are shown for (a) PMN-PT composition with x $\sim$ 0.12 (adopted from \cite{Colla2007}) and (b) for PMN-PT composition with x $\sim$ 0.295. Two cooling lines are plotted for two different cooling rates: 4 K/min and 0.5 K/min. The difference between these cooling lines is plotted as as a function of the DC electric field strength in the insets.}
    \label{fig:figure3}
\end{figure*}

To determine the concentration of each sample, we cooled them at a fixed rate ($\frac{dT}{dt} = 4$ K/min) from $T = 500$ K in the paraelectric region to $T \le 300$ K under concurrent application of a low strength AC electric field ($E_{AC} = 0.002$ kV/cm). No DC electric field bias was applied to the sample during the experiment. The real, linear dielectric susceptibility was measured during the entirety of the experiment, allowing us to plot it as a function of the sample temperature. After obtaining the temperature of peak dielectric susceptibility, $T_{max}$, we used a linear empirical fit that we performed on the data provided in \cite{Zekria05} for $x < 0.5$:

\begin{equation}
    x = \frac{T_{max} - 259}{517}
\end{equation}

From this, we found the approximate concentrations of the samples that we studied in this report. The concentrations of the samples are also marked on Fig.~\ref{fig:figure1}. 

\section{Results}
\subsection{Field-Cooling Regime}

The field cooling (FC) and field heating (FH) regime is defined by the application of a DC field at high temperature and ramping the temperature up and down with a fixed value of the DC field. Above a certain threshold DC electric field, PMN-PT will undergo a field-induced ferroelectric phase transition. The transition temperature was established by the polarization/depolarization current peaks. Our experimental investigations study how $T_c$ changes due to different applied temperature-field histories in the FC-FH regime. 

Fig.~\ref{fig:figure2} (a) shows a typical FC experimental protocol. The polarization current spikes associated with a macroscopic field-induced ferroelectric transition in the Field-Cooling and Heating regime are provided in Fig.~\ref{fig:figure2} (b). A singular polarization current spike is observed when the PMN-PT composition undergoes a transition into the ferroelectric phase. During the melting of the polarization, two polarization current spikes are observed, where one has a notably smaller amplitude but covers a broader temperature range. The other polarization current has a narrower temperature range and higher amplitude. Through this experiment, the threshold electric-field $E_{th}$ (minimum electric field require to observe an induced ferroelectric transition) was determined. Next, this experiment was repeated at different strengths from 0 to 0.4 kV/cm. In addition, it was performed for two different cooling rates $\frac{dT}{dt} = 0.5$ and $4$ K/min. From the results of this experiment, an electric field-temperature (T-E) phase diagram was constructed for each sample similar to Fig.~\ref{fig:figure3} (a).

\begin{figure*}[t]
    \centering
    \includegraphics[width=1.0\linewidth]{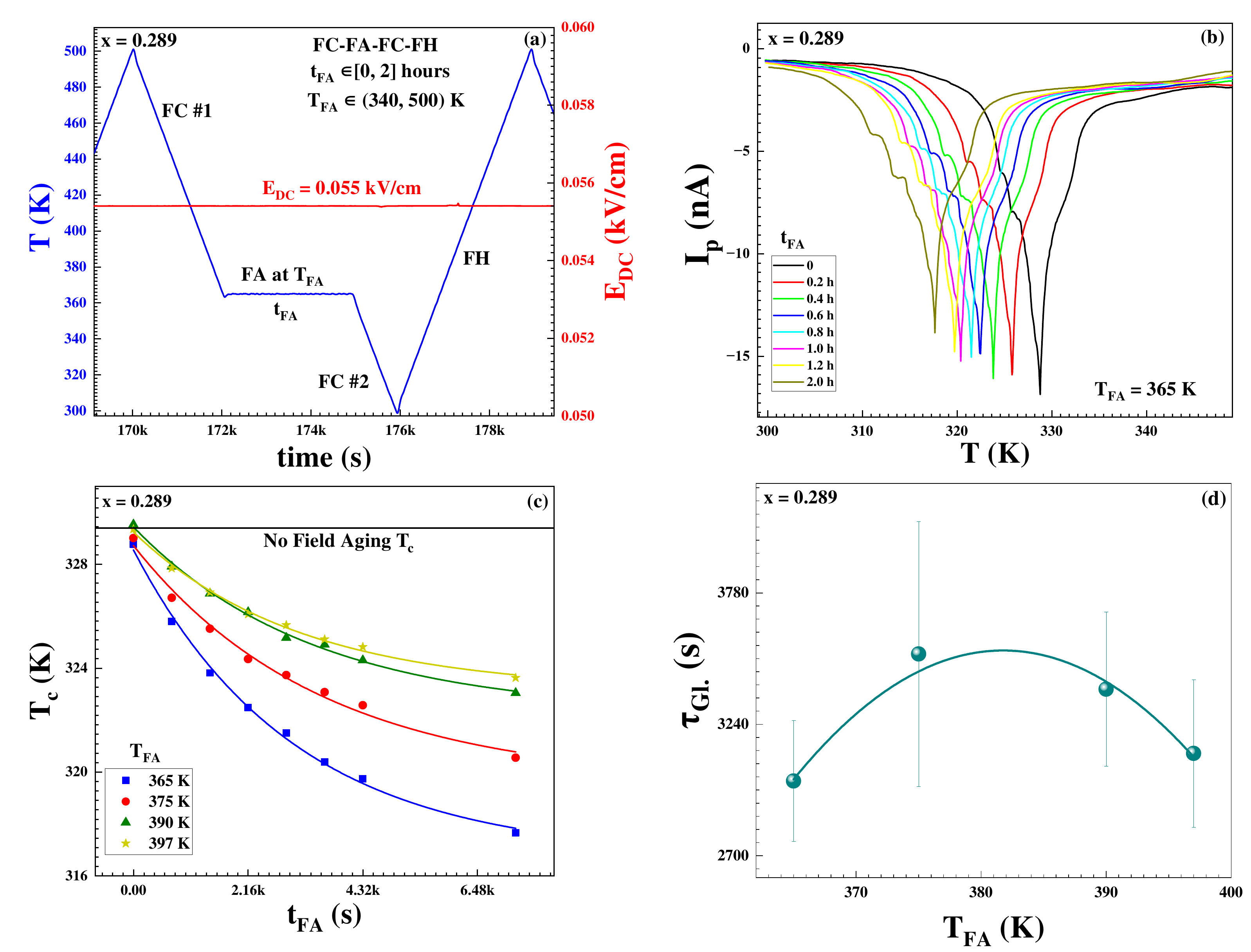}
    \caption{\justifying (a) Intermediate Field-Aging step protocol showing temperature and external electric field strength as functions of time. (b) Field-induced polarization current in FC\#2 plotted as a function against  different aging times ranging from 0 to 2 hours. (c) Characteristic curves of field-induced phase transition temperatures plotted as a function of the aging time for four different aging temperatures. (d) Characteristic time of transition temperature decay---extracted from exponential fits in panel (c)---plotted as function of $T_{FA}$.}
    \label{fig:figure4}
\end{figure*}

Fig.~\ref{fig:figure3} compares the $E$-$T$ phase diagrams for the field-induced phases in the $x = 0.295$ composition studied in this report to the $x = 0.12$ composition studied in \cite{Colla2007}. For the sample with $ x = 0.295$, we observed two steps of melting  of the ordered phase. In Fig.~\ref{fig:figure3} (b), the cooling line for $dT/dt = 0.5$ K/min is behind the cooling line for $dT/dt = 4$ K/min in the $x = 0.295$ sample. This is opposite of the behavior seen in the low-concentration sample $x = 0.12$, where the empirical cooling line for $dT/dt = 0.2$ K/min is ahead of the $dT/dt = 4$ K/min cooling line. The inset graphs of Fig.~\ref{fig:figure3} further quantify the differences between the transition temperatures for the different cooling rates. This difference is eliminated as the strength of the electric field is increased, asymptotically reaching zero. 

These results primarily indicate that field-induced phase transition dynamics in PMN-PT are strongly dependent on electric field and temperature (field-temperature) history. There is a physical difference between cooling at different rates---it is how long the sample spends in a non-ferroelectric state. Given that the endpoint temperatures were fixed in our protocols, a sample that is cooled at $dT/dt = 4$ K/min spends less time in a non-ferroelectric state compared to a sample that is cooled at $dT/dt = 0.5$ K/min. Hence, our results imply that more time-spent in a non-ferroelectric state can notably change induced ferroelectric phase transition dynamics.

\begin{figure*}[t!]
    \centering
    \includegraphics[width=0.905\textwidth]{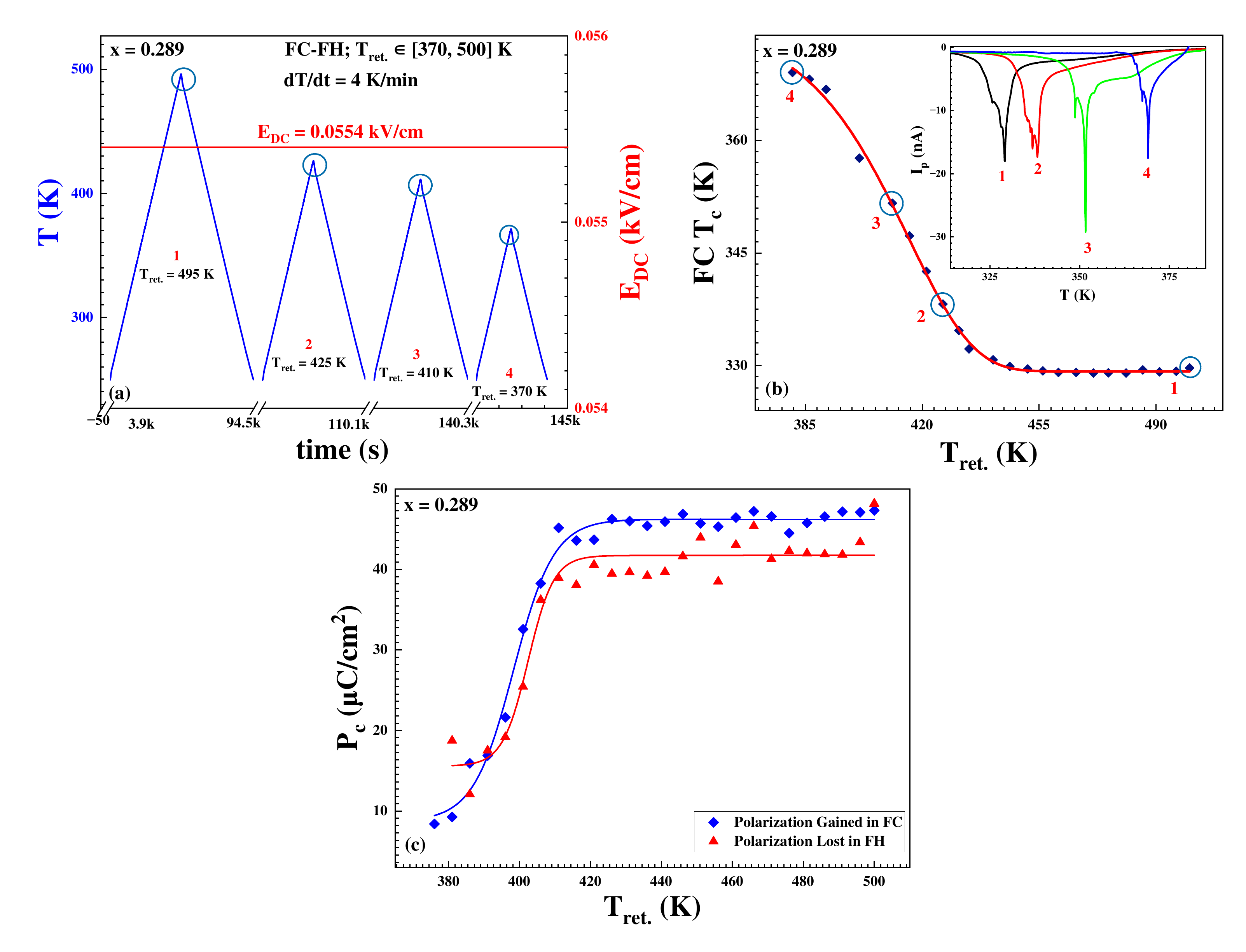}
    \caption{\justifying (a) Certain time slices of the different return point temperature experimental protocol are shown. (b) The empirical field-induced ferroelectric phase transition temperature, $T_c$, is plotted against the return point temperatures. Four return point temperatures are circled in both (a) and (b). The inset graph plots the induced polarization current for the four circled points against temperature. (c) The polarization gained (marked in blue) and lost (marked in red) in field-induced phase transitions is graphed as a function of return point temperature.}
    \label{fig:figure5}
\end{figure*}

Next, we examine the effect of an additional intermediate isothermal aging step on the temperature cycle. The temperature of this field-aging step ($T_{FA}$) and the time spent in the isothermal field-aging step ($t_{FA}$) were varied in this experiment. Our goal was to understand spending more time in which temperature region was responsible for the observed temperature lag between fast and slow cooling rates in the $E$-$T$ phase diagram. 

In Fig.~\ref{fig:figure4}, the protocol of the experiment (a), examples of raw data (b), and the results of significant reduction of the transition temperature $T_c$ (c) are shown. It is also shown that as the aging temperature is decreased to near the no-aging transition temperature, the rate at which $T_c$ is reduced notably increases. This result implies that aging in a temperature region below the two melting lines in the phase diagrams (see Fig.~\ref{fig:figure3}) significantly dampens the kinetics in field-cooled phase transitions. In addition, the observed temperature lag gets larger as $t_{FA}$ increases. It is probable that formation of glassy order in the field-aging step is responsible for this phenomena. Spending $t_{FA}$ in the aging step likely determines the rigidity of glassy order formed, while performing the aging at $T_{FA}$ could determine how quickly the glassy order is formed in the aging step. Fig.~\ref{fig:figure4} (d) potentially supports this claim as it shows the characteristic time, $\tau_{Gl.}$, of the change in induced transition temperature due to field-aging at different $T_{FA}$. This was obtained from the exponential fits provided in Fig.~\ref{fig:figure4} (c). If glassy ordering was responsible for this delay, $\tau_{Gl.}$ could represent a kinetic constant, capturing the time required for glassy-like effects to significantly quench system dynamics. The time-scale for such quenching is shown to be on the order of hours ($\sim$ 1000 seconds) and is strongly dependent on the aging temperature, $T_{FA}$. 

Our second investigation determined if cooling from different temperatures on the phase diagram significantly impacted $T_C$. The FC-FH cycles were performed with step-by-step reduction of the maximum annealing temperature as shown in Fig.~\ref{fig:figure5} (a)---denoted by $T_{ret.}$. Another interesting point is that the ratio of time spent in the ferroelectric state to the time spent in non-ferroelectric states increases as $T_{ret.}$ is reduced. Hence, this experiment investigated how the polarization lost/gained in the heating/cooling step varied across different temperature regions.

From Fig.~\ref{fig:figure5} (b), it is clear that the FC transition temperature rapidly increases with decreasing of annealing temperature $T_{ret.}$ from $440$ K and remains almost constant when $T_{ret.}$ is higher than $440$ K. The range of the temperature shift in $T_c$ is approximately $\Delta T \sim 40$ K, demonstrating a significant difference between field-induced phase transitions with different starting high-point temperatures. As seen in Fig.~\ref{fig:figure5} (c), the polarization gained and lost, $P_c$, remains relatively constant in $T_{ret.}$ from $420$ to $500$ K. Below $T \le 420$ K, $P_c$ drastically drops, following a similar logistic trend evident in Fig.~\ref{fig:figure5} (b). This indicates that not all the polarization is melted when field-heating to a temperature near the melting lines---the sample has retained a fraction of the saturated polarization. As indicated by the shift in $T_c$, it is likely that retained polarization can kinetically accelerate the induced phase transition. In addition, a temperature lag between where the sharp decrease in $P_c$ and sharp increase in $T_c$ happen is observed. This suggests that the retention of polar order may occur at the microscopic scale, potentially acting as a trigger for this phenomena. 

\begin{figure}[t]
    \centering
    \includegraphics[width=0.485\textwidth]{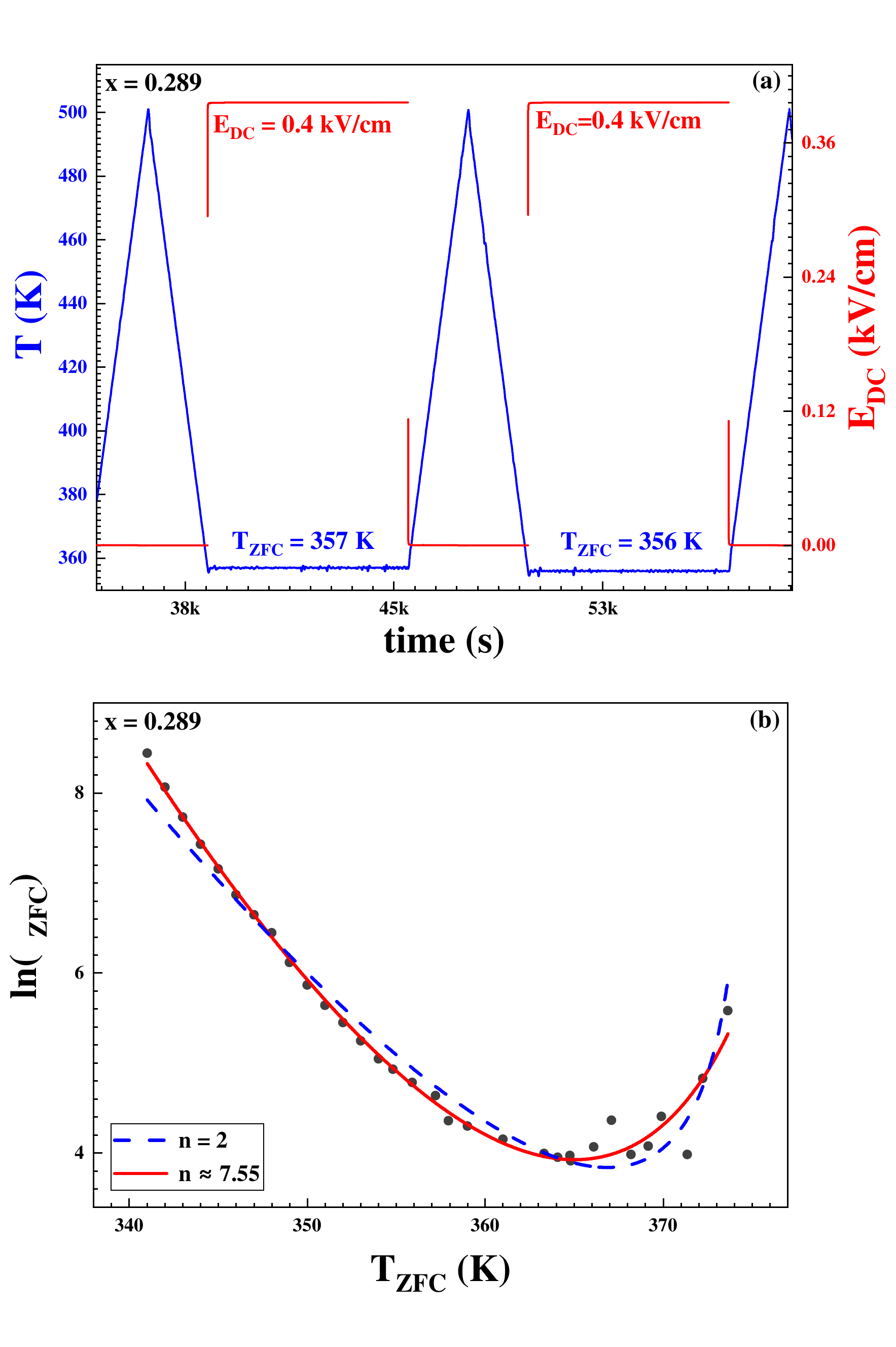}
    \caption{\justifying (a) The temperature-field protocol for the regular zero-field cooling/heating regime is shown as electric field and temperature plotted as functions of time.  $T_{ZFC}$ was varied between 340 and 375 K. (b) The natural logarithm of $\tau_{ZFC}$ is plotted as a function of $T_{ZFC}$. Two fits are plotted for different exponents, $n$, in Equation \ref{eq:equation2}.}
    \label{fig:figure6}
\end{figure}

\subsection{Zero Field-Cooling Regime}

The zero-field cooling (ZFC) regime is specified by cooling a sample in zero field to a temperature below the freezing line in Fig.~\ref{fig:figure3} and applying a DC electric field at that fixed temperature. The transition to the long-range induced ferroelectric state appears not instantly but after some delay time, $\tau_{ZFC}$, which depends on the temperature and value of the DC field \cite{Colla1995}. A set of ZFC experiments (analogous to the FC regime) was performed to study how $\tau_{ZFC}$ depends on the field-temperature history of the system.

Our first experiment focused on investigating $\tau_{ZFC}$ as a function of the isothermal relaxation temperature ($T_{ZFC}$) for a fixed value of applied $E_{DC}$ (see Fig.~\ref{fig:figure6} (a)). On Fig.~\ref{fig:figure6} (b), $\tau_{ZFC}$ rapidly increases for lower $T_{ZFC}$ when $T_{ZFC} \leq 365$ K. In this experimental protocol, $\tau_{ZFC}$ can also be postulated as a nucleation lag time as presented in \cite{Colla2014PRB}. In this context, the observed time-dependence indicates a deviation from the typical Arrhenius-like behavior---similar to the trend observed in compositions of PMN-PT at $x = 0.20$ \cite{Colla2014PRB}. The data was fit to the following equation:

\begin{equation}
    \ln(\frac{\tau}{\tau_0}) = \frac{T_1}{T} - n\ln{(1 - \frac{T}{T_C})}
    \label{eq:equation2}
\end{equation}

In this equation, $\tau_0$ and $T_1$ are fit parameters, while $n$ represents the rate of divergence of the nucleation lag time. The choice of this fit stems from the leading order approximation ($\Delta G(T) \sim [T - T_C]$) of the classical Arrhenius homogeneous nucleation theory prediction \cite{Fokin2006}:

\begin{figure}[t!]
    \centering
    \includegraphics[width=1.0\linewidth]{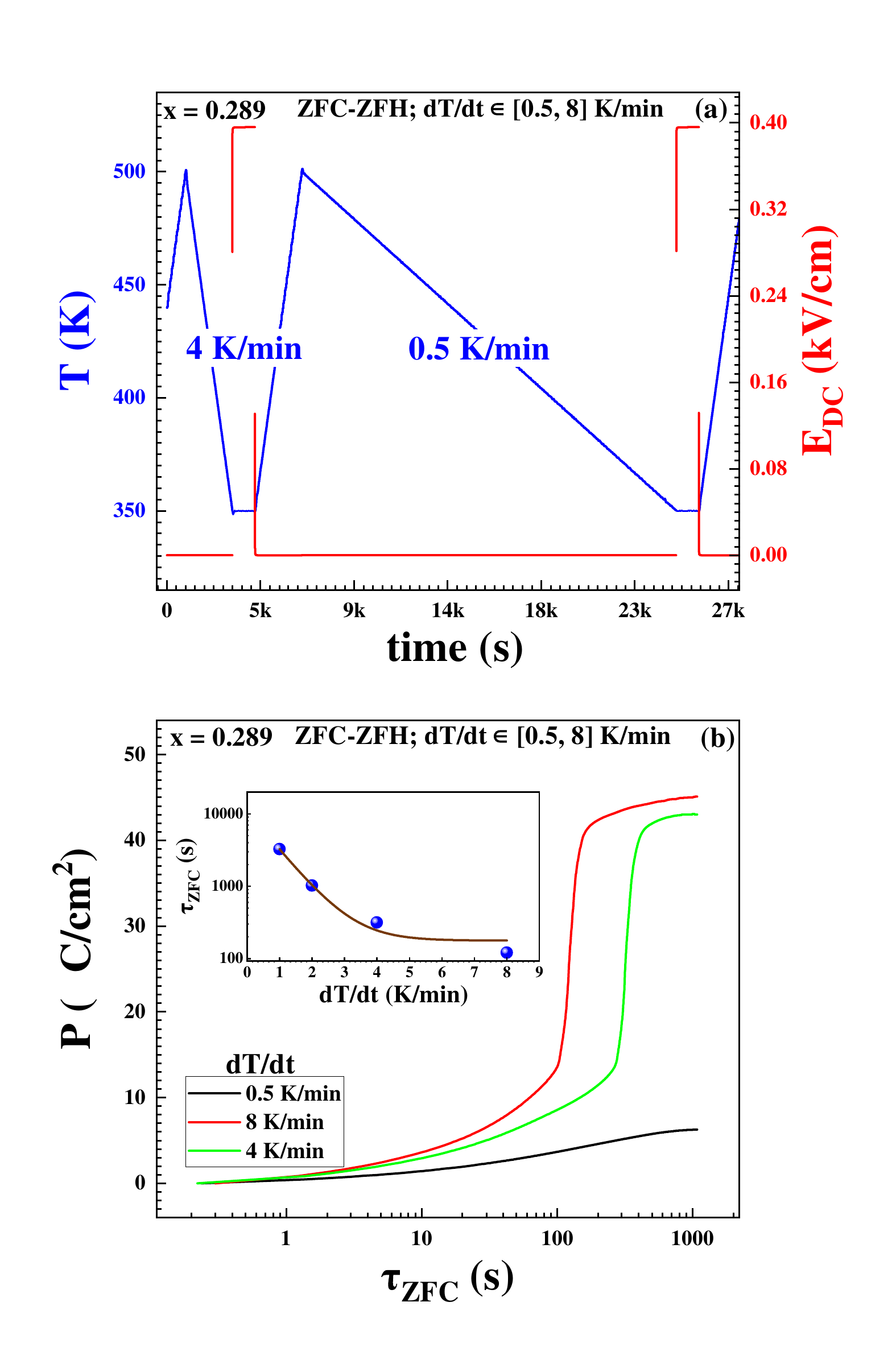}
    \caption{\justifying (a) The empirical protocol for performing zero-field cooling/heating is shown for two cooling rates: $dT/dt$ = 4 K/min and 0.5 K/min. (b) The polarization is plotted as a continuous function of time on a log base-10 scale during the ZFC-relaxation step for three different cooling rates: $dT/dt$ = 0.5, 4, and 8 K/min. The inset shows the dependence of $\tau_{ZFC}$ for all experimentally tested cooling rates.}
    \label{fig:figure7}
\end{figure}

\begin{figure*}[htbp!]
    \centering
    \includegraphics[width=1.0\linewidth]{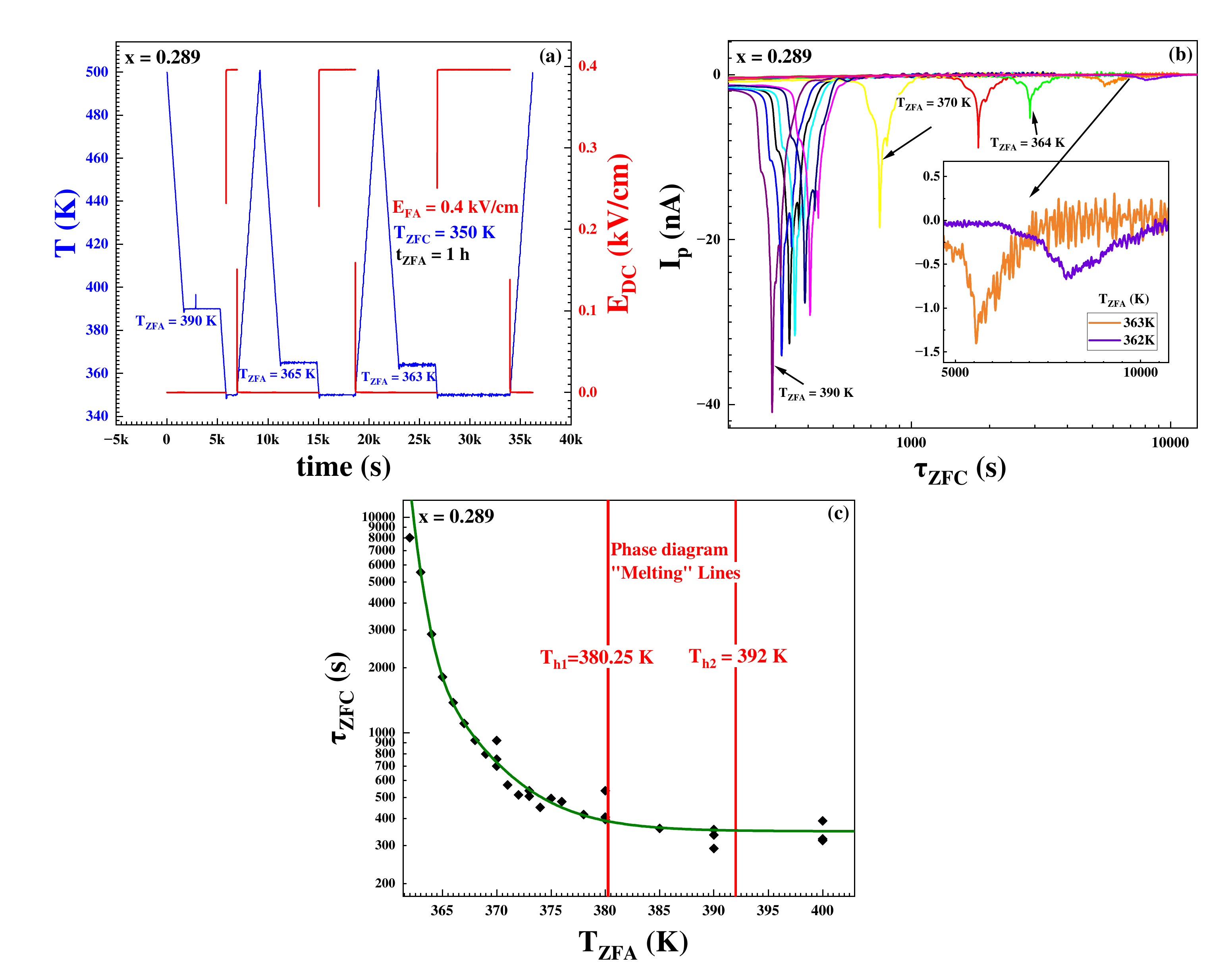}
    \caption{\justifying (a) The intermediate zero-field aging step experimental protocol is plotted as temperature and DC electric field strength as functions of time. The zero-field aging duration, $t_{ZFA}$, was fixed at an hour and the ZFC-relaxation temperature, $T_{ZFC} = 350$ K, was fixed for all experiments. (b) The polarization current is plotted against the time delay until ferroelectricity, $\tau_{ZFC}$ for varying $T_{ZFA}$. A zoomed-in inset plot is provided for polarization currents with $T_{ZFA} = 362$ and $363$ K . (c) $\tau_{ZFC}$ is plotted against $T_{ZFA}$.}
    \label{fig:figure8}
\end{figure*}

\begin{equation}
    \tau \propto \frac{\exp{\frac{B}{kT}}}{\Delta G(T)^2}
    \label{eq:equation3}
\end{equation}

In Equation \ref{eq:equation3}, $B$ represents the Arrhenius Barrier height, $k$ is the Boltzmann constant, and $\Delta G(T)$ is the bulk free-energy difference between the ordered and disordered phases \cite{Colla2014PRB}. If the nucleation process is described by Equation \ref{eq:equation3}, then n should be equal to 2 in Equation \ref{eq:equation2}.

Our result indicates that the nucleation lag time only qualitatively follows the temperature dependence in Equation \ref{eq:equation3}. This is in accordance with prior work on PMN-PT compositions with $x = 0.20$ \cite{Colla2014PRB}.  The $n = 2$ curve fails to fit to the low-temperature data as the rate of divergence in $\tau_{ZFC}$ is much steeper than expected. When $n$ was made an adjustable fitting parameter, the best fit found was for $n \sim 7.5$. This exponent is larger than any best-fit exponent found in PMN-PT compositions with $x = 0.12$ and $0.20$ \cite{Colla2014PRB}. This is preliminary evidence of enhanced slowing of induced phase-transition dynamics in compositions of PMN-PT near the MPB. The system exhibits also rapid $\tau_{ZFC}$ divergence for $T_{ZFC} > 365$ K. This is within expectations as a non-ferroelectric state can provide an increase in entropy that outweighs the energetically-favorable ferroelectric state.

\begin{figure*}[htbp!]
    \centering
    \includegraphics[width=1.0\linewidth]{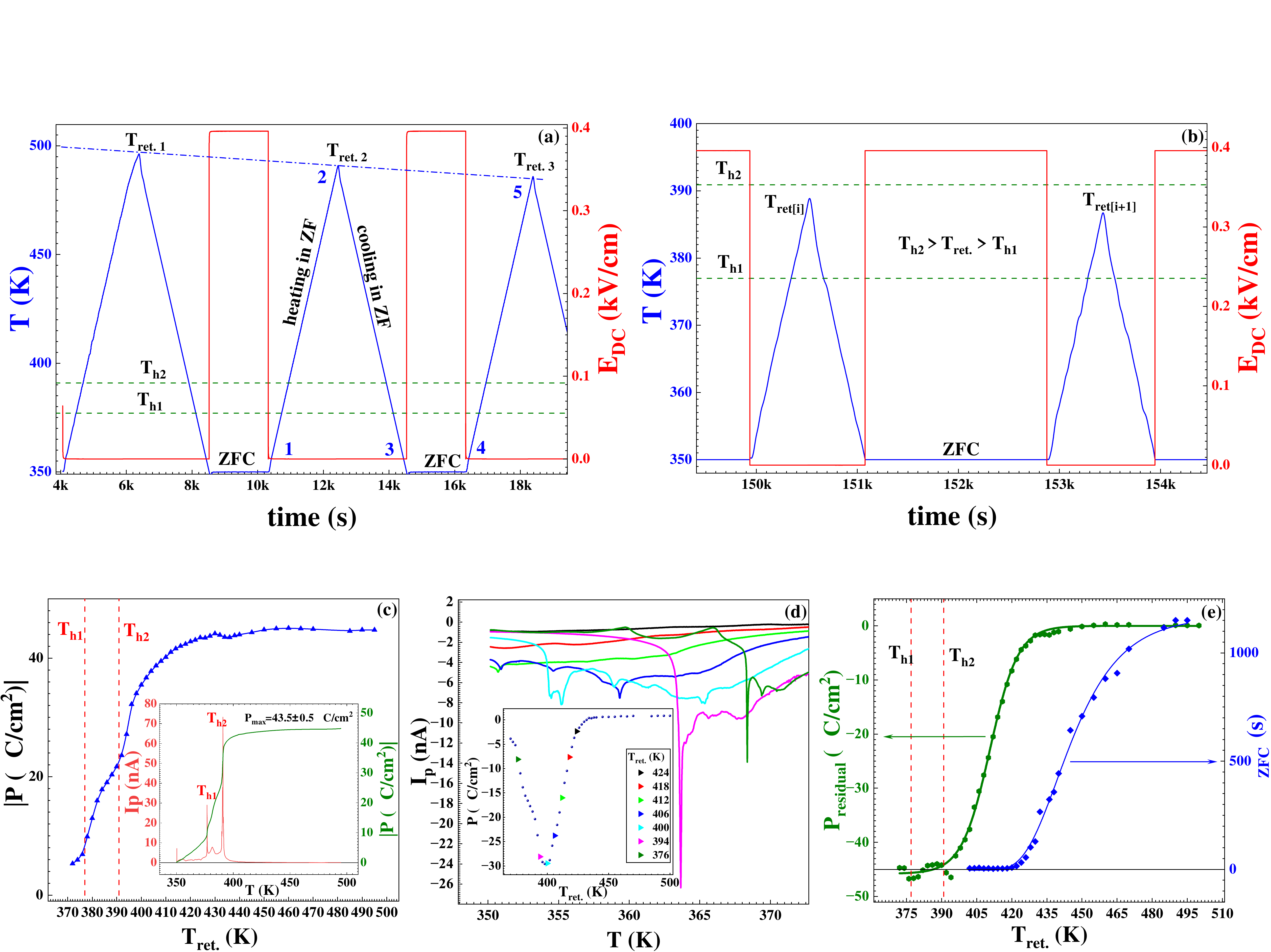}
    \caption{\justifying ZFC Different Return Point Temperature experimental protocol is shown for $T_{ret.} \gg T_{h1}$ \& $T_{h2}$ (a) and $T_{h1} < T_{ret.} < T_{h2}$ (b). (c) Melted Polarization in heating step is plotted as a function of the return point temperature. (d) Polarization Currents in the no-field cooling step are plotted for different return point temperatures. In the inset graph, the polarization gained in this step is plotted for all return point temperatures. (e) The residual polarization after every ZFC cycle and ZFC relaxation time, $\tau_{ZFC}$, are displayed for all return point temperatures.}
    \label{fig:figure9}
\end{figure*}

Next, the ZFC experiment was repeated with varying cooling rates while approaching the ZFC temperature. Fig.\ref{fig:figure7} (b) shows the change in system polarization as a function of time for varying ZFC cooling rates. The polarization was calculated as a function of time from the start of the ZFC-relaxation step to every data-acquired point in time. The delay time, $\tau_{ZFC}$, significantly increased as the cooling rate decreased. In all ZFC experiments, we observe a polarization creep to a critical polarization (typically 1/3 of the saturated polarization), which is followed by an avalanche-like phase transition into the ferroelectric state. As seen in the inset of Fig.\ref{fig:figure7} (b), this polarization creep rate is strongly dependent on the cooling rate. Approaching $T_{ZFC}$ at a slower cooling rate also slowed the polarization creep to the critical polarization. The natural tendency to form short-range glassy order has been previously argued in low-concentration compositions of PMN-PT \cite{Colla2007, Colla2008}. Hence, it is likely the system experiences an enhanced competition between the formation of glassy and ferroelectric order in slower-cooling rate experiments, where more time is spent in a non-ferroelectric state.

A Zero-Field Intermediary Aging Step experiment was performed to study this possibility, testing if aging in a particular region of the FC-FH phase diagram significantly impacted $\tau_{ZFC}$. The protocol of this experiment is shown in Fig.~\ref{fig:figure8} (a). There are two isothermal aging steps. The first aging step was done with no external electric field applied. The second aging step is the transition inducing step---similar to the standard ZFC experiment described above. The variable parameter in this experiment was the temperature of first zero field aging, $T_{ZFA}$ (refer to Fig.~\ref{fig:figure8} (a)).

\begin{figure*}[t!]
    \centering
    \includegraphics[width=1.0\textwidth]{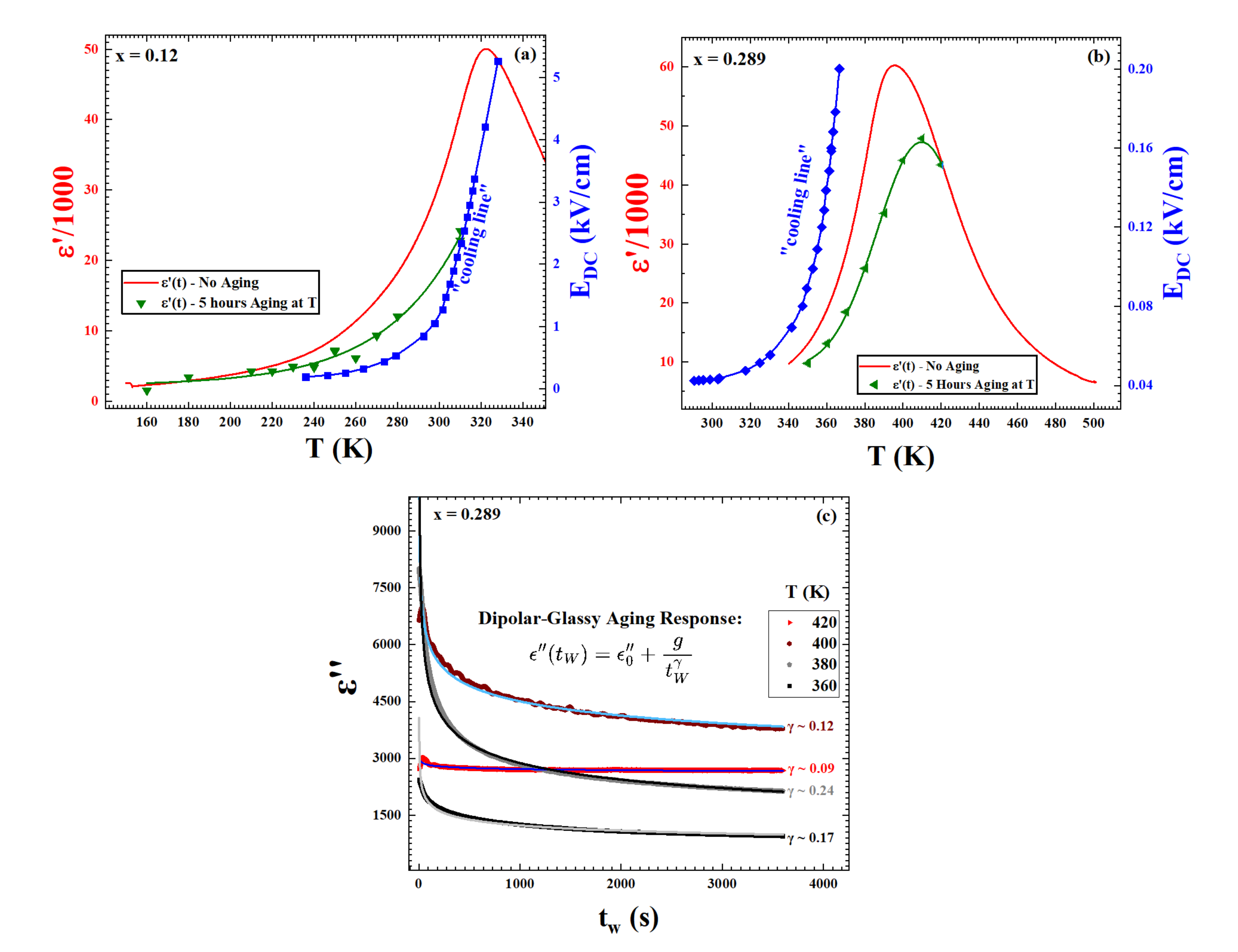}
    \caption{\justifying Real component of the scaled dielectric susceptibility, $\epsilon'(T)$, is plotted for (a) no-aging continuous measurement and (b) after 5 hours of aging at fixed $T$. ``Cooling Line" refers to the field-induced ferroelectric transition temperatures on Fig.~\ref{fig:figure3}. (c) shows the dielectric loss response, $\epsilon''(T)$, during the aging steps for 1 hour at $T$ = $420$, $400$, $380$, $360$ K. These responses were fitted with Equation \ref{eq:equation4}. The $x = 0.12$ data was adopted from \cite{Colla2007} and \cite{Chao2006}.}
    \label{fig:figure10}
\end{figure*} 

 The result of this experiment clearly show significant increasing of the delay time, $\tau_{ZFC}$, starting below the temperatures of the two melting lines. Raw data of the polarization current spikes observed in the ZFC relaxation step is shown in Fig.~\ref{fig:figure8} (b), and final results of this experiments are presented in Fig.~\ref{fig:figure8} (c). This result is coherent with the different cooling rate experiment and shows that spending more time in the temperature region below the melting temperatures (representative of the relaxor state) fundamentally impacts the transition to the long-range ordered phase in the ZFC regime. The effect of zero-field aging amplifies when $T_{ZFA}$ is close to the ZFC-relaxation step temperature, $T_{ZFC}$.

\begin{figure*}[t!]
    \centering
     \includegraphics[width=1.0\textwidth]{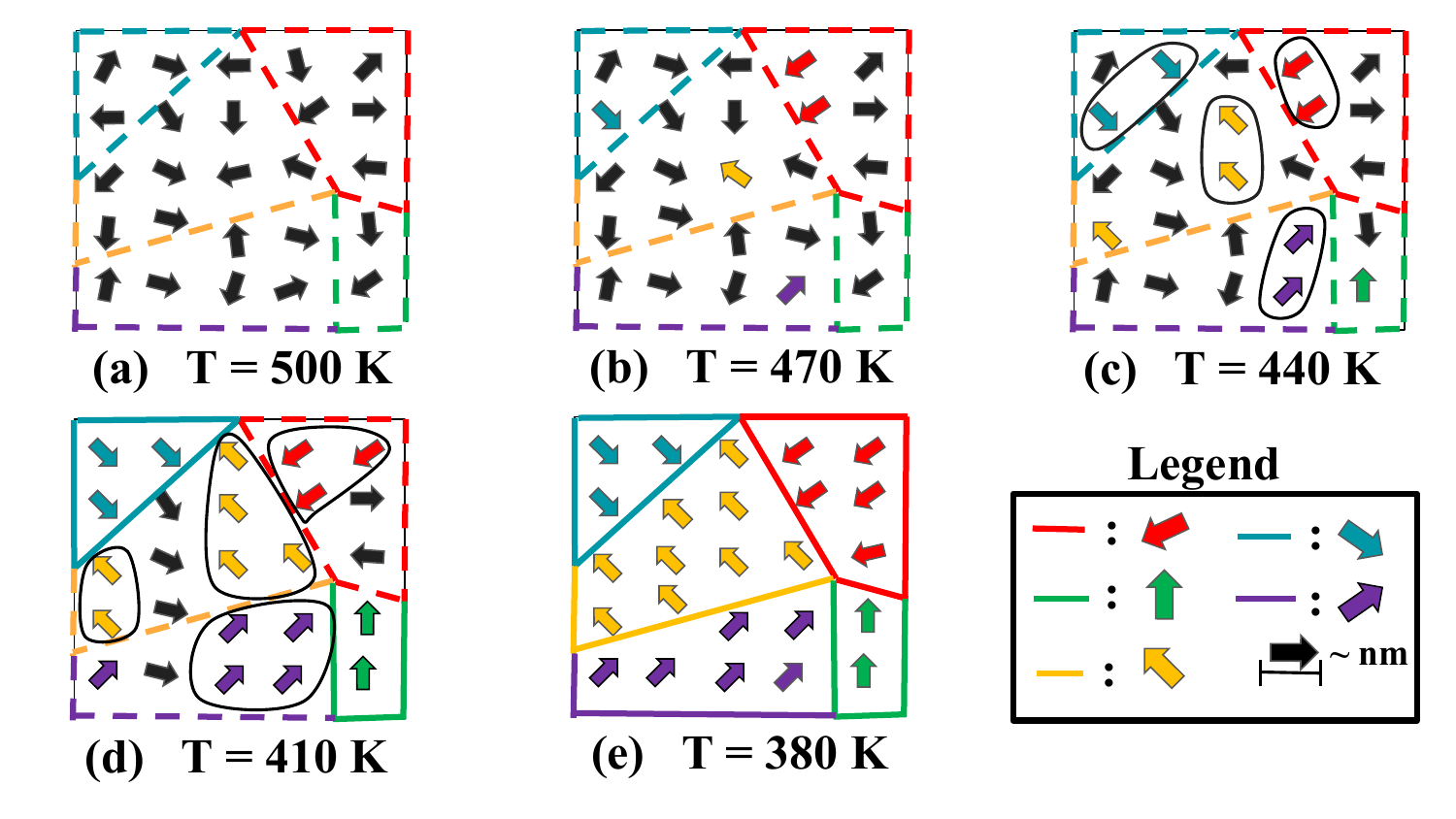}
    \caption{\justifying A visual schematic of the polarization retention phenomenon observed in the different $T_{ret.}$ experiment is provided. The thick colored dashed lines represent boundaries of prior ferroelectric domains, where the polarization direction of the domain is provided in the legend. These dashed lines become filled lines when an entire domain has not been depolarized. Each arrow represents a short-range polar ordered (SRPO) region with typical nanometer length scale and is colored when it is direction is left invariant upon depolarization. Snapshots of the internal distribution of SRPO is shown at different return point temperatures from (a) $T = 500$ to (e) $T = 380$ K. To illustrate macroscopic retention, a circle is drawn on clusters of SRPO that are not melted upon entering the non-ferroelectric state.}
    \label{fig:figure11}
\end{figure*} 

Finally, a ZFC different return point temperature experiment was performed to determine the cooling/heating behavior from different temperatures of ZFC phase transitions. Fig.~\ref{fig:figure9} (a) and (b) shows the experimental protocols for two temperature regions: above and between the melting lines. Starting at point 1 on Fig.~\ref{fig:figure9} (a), the sample is fully polarized from the previous ZFC relaxation step. From the observed depolarization current, the polarization lost in each heating step can be approximated. It is important to note that the maximum obtainable polarization can be estimated when the sample is heated up to $T_{ret.} = 500$ K. Fig.~\ref{fig:figure9} (c) shows the amount of polarization lost after heating the sample to $T_{ret.}$, with the inset focusing on the depolarization record for $T_{ret.} = 500$ K. It is clear that the sample does not lose all polarization even when it is heated over both melting temperatures ($T_{h1}$ and $T_{h2}$ on the figure). After heating to $T_{ret.}$ at point 2, the next step is to cool back to the ZFC relaxation temperature, $T_{ZFC} = 350$ K (point 3). Fig.\ref{fig:figure9} (d) shows the polarization currents obtained in the no-field cooling steps. It is evident that ferroelectric self-ordering is occurring without the application of an external electric field. This ferroelectric ordering may be coerced due to the unmelted polarization. The inset graph shows the polarization gained during the cooling steps from different $T_{ret.}$. The polarization gained in the step depends on $T_{ret.}$, where the maximum polarization change is located at $T_{ret.} = 400$ K. 

The last step of the cycle is performing the ZFC-relaxation step (from point 3 to point 4). Given that not all polarization is melted away in the heating step and some polarization is gained in the cooling step, the ZFC-relaxation step may start some with non-zero polarization. Hence, Fig.\ref{fig:figure9} (e) shows the residual polarization of the sample prior to performing the ZFC-relaxation step (green line). The residual polarization was calculated by subtracting the polarization lost in the heating step from the maximum system polarization and then adding this quantity to the polarization gained in the cooling step. Fig.\ref{fig:figure9} (e) also shows the ZFC delay time as a function $T_{ret.}$ (blue line). This data could only be obtained up to $T_{ret.} = 400$ K due to our limited time-resolution. Given that $\tau_{ZFC}$ decreases rapidly at much higher $T_{ret.}$, it is clear that even a small amount of residual polarization can significantly accelerate the transition to the ferroelectric ordered phase during the ZFC step. Specifically, the system does not require a macroscopic polarization to generate this kinetic acceleration, hinting that this phenomena's origin lies within the short-range polar order. 

\section{Discussion}
The key interpretation of our experiments is that the dynamics of induced ferroelectric transitions in PMN–PT compositions near the morphotropic phase boundary (MPB) exhibit a field–temperature history dependence that differs significantly from the history dependence observed in compositions far below the MPB. In our experiments, we controlled the history profile of these compositions in two regimes: the time spent in non‑ferroelectric states and the time spent in ferroelectric states. Accordingly, we divide this discussion into these two regimes and describe potential mechanisms responsible for the observed impacts on field‑induced ferroelectric phase transitions.

\subsection{Formation of Glassy Order in Non-Ferroelectric State}
The FC-FH phase diagrams show that, for compositions near the MPB (see Fig.~\ref{fig:figure3} (b)), increasing the time spent in the non-ferroelectric state lowers the field-induced transition temperature. This trend is further reinforced by the aging experiments, which demonstrate that isothermal aging below the melting temperatures also reduces $T_c$ (see Fig.~\ref{fig:figure4} (c)).

This behavior contrasts with low-concentration compositions ($x = 0.12$), where slower cooling increases $T_C$. The key questions is therefore: why does additional time in the non-ferroelectric state suppress the transition near the MPB, despite the higher ferroelectric component concentration?

Fig.~\ref{fig:figure10} provides insight. The dielectric susceptibility  $\epsilon'(T)$ shows that aging occurs in different temperature ranges depending on composition: below the transition line for $x = 0.12$, but well above it for $x \approx 0.289$. This indicates that the underlying aging mechanism differs between these regimes. Multiple mechanisms---including the formation of glassy order---could be responsible for the aging of $\epsilon'$. However, different aging mechanisms have different functional forms of the dielectric loss response, $\epsilon''$, during the aging: for glassy-like freezing, an inverse-power law of the dielectric loss response is observed \cite{Chao2006}.

\begin{equation}
    \epsilon''(t_{W}) = \epsilon''_0 + \frac{g}{t_{W}^{\gamma}}
    \label{eq:equation4}
\end{equation}

In Equation \ref{eq:equation4}, $t_{W}$ represents the waiting time of the aging at a fixed temperature $T$, and $\gamma$ is an exponent that captures how quickly $\epsilon''$ decays. Fig.\ref{fig:figure10} (c) shows that the dielectric loss response in $x = 0.289$ (at temperatures above the melting transition line) fits well with the phenomenological reference Equation \ref{eq:equation4}. Additionally, $\gamma$ is roughly in the interval from $0.09$ to $0.25$. This agrees with the exponent found in previously published experimental literature on reentrant ferromagnet \ce{Cu_{0.20}Co_{0.80}Cl2-FeCl3}, Heisenberg-like spin-glass \ce{CdCr_{1.7}In_{0.3}S4}, and low-concentrations of PMN-PT where $\gamma \sim 0.07$ to $0.20$ in these systems \cite{Suzuki2004, Dupuis2001, Chao2006}. While this is not direct confirmation, the agreement with prior literature and current fit strongly suggests that the aging mechanism in $x = 0.289$ responsible above the melting transition is glassy-like in nature. Special aging experiments are required to directly verify this. 

This interpretation naturally explains the contrasting cooling-rate dependence. For $x = 0.12$, minimal glassy order forms above $T_C$, allowing polarization grow continuously once symmetry is broken leading to higher $T_C$ for slower cooling. In contrast, near the MPB ($x\approx 0.289$), glassy order develops in the non-ferroelectric state and inhibits the transition.

 This can be understood in terms of the free-energy landscape. It has been previously argued that the presence of random, anisotropic quenching fields could be responsible for a relaxor state---a source to prevent the formation of long-range ferroelectric order \cite{Vugmeister06}, \cite{Pirc99}. In the presense of random anisotropic fields, the system likely exhibits a rugged landscape with many metastable minima. \cite{Bray1980, Drossel2004}. During slow cooling or isothermal aging, the system can become trapped in glassy local minima. Escaping these states requires additional activation energy, delaying the onset of long-range ferroelectric order. Physically, the applied field must first mobilize locally frozen polar regions before macroscopic ordering can occur. In rapid cooling, the landscape evolves too rapidly for such trapping, suppressing glassy formation and eliminating the delay.

This framework is consistent with the ZFC aging results (refer to Fig.~\ref{fig:figure8} (c)), which show enhanced delay when more time is spent in the non-ferroelectric temperature region. It also explains why the cooling-rate dependence weakens at higher electric fields: stronger fields favor alignment with the external field over local quenched disorder, suppressing glassy behavior (see Fig.~\ref{fig:figure3}).

An open question is why glassy ordering appears at higher temperatures near the MPB. A plausible explanation is enhanced structural frustration. Near the MPB, competing tetragonal (PT-like) and rhombohedral (PMN-like) tendencies may generate additional disorder and anisotropic fields. This competition is reflected in the poorly defined phase boundaries reported in the literature \cite{Kim2022, Guo2003}. Given that the source of anisotropic quenching fields is compositional in nature \cite{Pirc99, Westphal92, Stock04}, enhanced structural frustration can lead to enhanced anisotropic quenching, promoting glassy behavior at elevated temperatures in these compositions.

Finally, a second inconsistency is that this mechanism should weaken as $x \rightarrow 0.4$, where the system behaves as a conventional ferroelectric. Identifying the concentration at which this crossover occurs remains an open problem. Further investigations into samples $0.3 < x < 0.4$---using a similar set of experimental protocols---are currently being performed to study this point in greater depth.

\subsection{Kinetic Acceleration of Ferroelectric Ordering}
In addition to the history dependence in the non-ferroelectric regime, our results reveal a distinct effect within the ferroelectric regime: prior field–temperature history accelerates subsequent ferroelectric ordering.

Specifically, the FC different return-point temperature experiments show that annealing near the empirical cooling line leads to field-induced transitions at higher temperatures. In the ZFC regime, the same history reduces the delay time 
$\tau_{ZFC}$, indicating faster transition kinetics. Furthermore, the system exhibits a memory effect: remnants of prior ferroelectric domains persist and can facilitate reordering, even enabling partial self-organization in zero external field.

A natural interpretation is a repoling-like mechanism involving SRPO regions, similar to behavior reported in other relaxor-ferroelectrics \cite{Xu2005, Granzow2002}. The key idea is that the depolarization of the samples is incomplete: although macroscopic polarization may vanish, a fraction of SRPO retains its prior orientation and acts as a seed for subsequent ordering (refer to Fig.~\ref{fig:figure9}). This mechanism is illustrated schematically in Fig.~\ref{fig:figure11}. At high return temperatures ($T_{ret.} \sim 500 $ K), all polarization memory is erased and SRPO orientations are randomized. As $T_{ret.}$ decreases, a subset of SRPO regions remains aligned with prior ferroelectric domains. These retained regions serve as local nucleation centers: they can couple to neighboring depolarized regions and promote cooperative realignment toward the previous domain orientation. 

With further reduction of $T_{ret.}$, this effect becomes collective. Clusters of retained SRPO emerge and grow, eventually leading to partially unmelted domains. Once such domains are present, the system no longer needs to nucleate ferroelectric order from a fully disordered state. Instead, it evolves from a partially ordered configuration, significantly lowering kinetic barriers. This picture is strongly supported by the return-point experiments. Even when the residual macroscopic polarization is zero, the delay time $\tau_{ZFC}$ decreases by orders of magnitude (refer to Fig,~\ref{fig:figure9}). This demonstrates that kinetic acceleration does not require macroscopic polarization, but is instead governed by mesoscopic SRPO.

As $T_{ret.}$ decreases further and macroscopic domains persist, the polarization changes during thermal cycling diminish. This is consistent with the observed reduction in both polarization gain and loss, as well as the saturation of residual polarization. In this regime, the system is effectively already ordered, and subsequent transitions become trivial. Overall, these results indicate that the kinetics of ferroelectric ordering near the MPB are controlled not only by thermodynamic conditions, but also by the degree of retained short-range order. Prior history determines whether the system must overcome a nucleation barrier or can instead grow from pre-existing aligned regions. This provides a unified explanation for the observed memory effects, accelerated ZFC dynamics, and field-free self-organization.

\section{Conclusion and Outlook}
In this work, we investigated the dynamics of field-induced ferroelectric phase transitions in PMN-PT compositions near the morphotropic phase boundary and their dependence on electric-field–temperature history.

We identify two distinct and competing effects. First, field-aging in the non-ferroelectric regime suppresses transition kinetics, leading to delayed field-induced ordering. Second, prior field–temperature cycling enhances kinetics, accelerating subsequent transitions. Together, these results demonstrate that phase-transition dynamics near the MPB are governed not only by instantaneous thermodynamic conditions, but also by the system’s history. These behaviors are naturally understood in terms of competing mechanisms: the formation of glassy order in the non-ferroelectric state and the retention of short-range polar order that facilitates subsequent ordering. This framework provides a unified interpretation of the observed kinetic slowing, acceleration, and memory effects.

These findings raise several open questions relevant to relaxor-ferroelectrics and disordered systems more broadly. How do different length scales of polar order interact and compete? What role does structural frustration play in stabilizing or inhibiting long-range order? To what extent are the observed kinetic effects universal across disordered ferroic systems? Addressing these questions will require further experimental and theoretical work. In particular, direct probes of short-range order (e.g., diffuse scattering) and systematic studies across compositions $0.3 < x < 0.4$ will be essential to determine the evolution from relaxor to conventional ferroelectric behavior.

\begin{acknowledgments}
We would like to thank students from the Fall 2025 semester of Physics 403: Modern Experimental Physics Laboratory course at the University of Illinois at Urbana-Champaign (UIUC), who worked on relevant data-collection of the final results. We would also like to thank ChatGPT and Copilot AI tools for their assistance in proof-reading, and providing edits/suggestions that enhanced the readability of this research paper. This work was sponsored by the Department of Physics at UIUC.
\end{acknowledgments}

\bibliography{references}

\end{document}